\documentclass[conference, hidelinks]{IEEEtran}
\PassOptionsToPackage{bookmarks={false}}{hyperref}
\IEEEoverridecommandlockouts
\usepackage{cite}
\usepackage{amsmath,amssymb,amsfonts}
\usepackage{dsfont}
\usepackage{algorithmic}
\usepackage{graphicx}
\usepackage{textcomp}
\def\BibTeX{{\rm B\kern-.05em{\sc i\kern-.025em b}\kern-.08em
    T\kern-.1667em\lower.7ex\hbox{E}\kern-.125emX}}
    
\ifCLASSOPTIONcompsoc
  \usepackage[caption=false,font=footnotesize,labelfont=sf,textfont=sf]{subfig}
\else
  \usepackage[caption=false,font=footnotesize]{subfig}
\fi

\usepackage{array}   % for \newcolumntype macro
\newcolumntype{C}{>{$}c<{$}} % math-mode version of "c" column type

\usepackage{xcolor}

\begin{document}

\title{ABE-Cities: An Attribute-Based Encryption \\ System for Smart Cities
%{\footnotesize \textsuperscript{*}Note: Sub-titles are not captured in Xplore and should not be used}
%\thanks{Identify applicable funding agency here. If none, delete this.}
}

\author{\IEEEauthorblockN{Marco Rasori\textsuperscript{*}\thanks{\noindent\textsuperscript{*} Also affiliated with University of Pisa, Dept. of Information Engineering.}\\}
\IEEEauthorblockA{\textit{DINFO} \\
\textit{University of Florence}\\
Florence, Italy \\
marco.rasori@unifi.it}
\and
\IEEEauthorblockN{Pericle Perazzo}
\IEEEauthorblockA{\textit{Department of Information Engineering} \\
\textit{University of Pisa}\\
Pisa, Italy \\
pericle.perazzo@iet.unipi.it}
\and
\IEEEauthorblockN{Gianluca Dini}
\IEEEauthorblockA{\textit{Department of Information Engineering} \\
\textit{University of Pisa}\\
Pisa, Italy \\
gianluca.dini@iet.unipi.it}
}

\maketitle

\begin{abstract}
In the near future, a technological revolution will involve our cities, where a variety of smart services based on the Internet of Things will be developed to facilitate the needs of the citizens.
Sensing devices are already being deployed in urban environments, and they will generate huge amounts of data.
Such data are typically outsourced to some cloud storage because this lowers capital and operating expenses and guarantees high availability.
%However, cloud storage may suffer of data breach vulnerabilities, as the recent Meltdown and Spectre flaws made evident, or it may have incentives to release stored data to unauthorized entities.
However, cloud storage may have incentives to release stored data to unauthorized entities.
% CONTRIBUTION:
In this work we present ABE-Cities, an encryption scheme for urban sensing which solves the above problems while ensuring fine-grained access control on data by means of Attribute-Based Encryption (ABE).
Basically, ABE-Cities encrypts data before storing it in the cloud and provides users with keys able to decrypt only those portions of data the user is authorized to access.
In ABE-Cities, the sensing devices perform only lightweight symmetric cryptography operations, thus they can also be resource-constrained.
ABE-Cities provides planned expiration of keys, as well as their unplanned revocation.
We propose methods to make the key revocation efficient, and we show by simulations the overall efficiency of ABE-Cities.

\end{abstract}

\begin{IEEEkeywords}
Smart City, Urban Sensing, Attribute-Based Encryption, Internet of Things, Data-Centric Security
\end{IEEEkeywords}

\IEEEoverridecommandlockouts
\IEEEpubid{\begin{minipage}{\textwidth}\ \\[2.5cm]
		\copyright \copyright 2018 IEEE. Personal use of this material is permitted. Permission from IEEE must be obtained for all other uses, in any current or future media, including reprinting/republishing this material for advertising or promotional purposes, creating new collective works, for resale or redistribution to servers or lists, or reuse of any copyrighted component of this work in other works.
\end{minipage}} 

%\IEEEpubidadjcol
\section{Introduction}

Smart cities offer digital services to facilitate the needs of their citizens and improve their quality of life.
Such services span across all sectors of society including health, logistics, and mobility.
Services often capitalize on the Internet of Things (IoT) as enabling technology.
For example, cameras could be deployed in the city so that citizens can monitor traffic and detect possible congestions on the usual route from home to work.
In general, smart devices underlying these services produce a large amount of data which can be outsourced to a Cloud Storage Service (CSS).
This is usually preferred to an in-house solution because it reduces costs while providing high availability.

In many cases, sensed data includes sensitive or valuable information which is intended to be read only by a set of authorized users.
Unfortunately, as soon as data lands on the cloud, we lose any control on it and we have to totally trust the CSS.
Cloud service providers may have some incentive to release stored information to others~\cite{di2007over,coppolino2017cloud}.
It follows that the CSS is generally assumed \emph{honest-but-curious}, meaning that it trustworthy carries out data distribution and elaboration functions but is interested in accessing the resource contents.

A possible solution to this problem is to store an encrypted version of the data on the CSS.
An efficient way to do this while ensuring fine-grained access control on data is to use the \emph{Attribute-Based Encryption} (ABE)~\cite{goyal2006attribute} technology.
ABE allows us to label a ciphertext with a set of \emph{attributes}.
A \emph{Trusted Third Party} (TTP) generates \emph{decryption keys} and provides them to users.
Decryption keys embed an access policy over the attributes, and they are capable of decrypting a ciphertext only if the set of attributes labelling the ciphertext matches the access policy.
%Apart from preventing service providers being curious, encrypting data stored on the CSS also limits the impact of data breaches on the cloud servers, which the recent Meltdown~\cite{lipp2018meltdown} and Spectre~\cite{kocher2018spectre} flaws made evident to be possible.

% CONTRIBUTION:
In this paper we present ABE-Cities, an encryption system for urban sensing in a smart city based on Attribute-Based Encryption.
ABE-Cities allows for fine-grained access control over the encrypted data stored in the CSS, and it is secure against multiple adversary models: a honest-but-curious cloud service, external adversaries capable of eavesdropping traffic and compromising sensing devices, and colluding users wanting to gain more authorizations illegally.
Sensing devices perform only lightweight symmetric cryptography operations.
Therefore, ABE-Cities can employ also resource-constrained sensing devices, which makes it suitable for a broader set of IoT applications for smart cities.
Our system also provides mechanisms to enforce planned expiration of decryption keys, as well as their unplanned revocation.
We propose methods to make the unplanned revocation efficient from the point of view of the cloud service.
We show by simulations the effectiveness of such improvements.

The paper is organized as follows: in Section~\ref{relatedwork} we compare with related work.
In Section \ref{background} we introduce ABE and other techniques that we use in our system.
In Section~\ref{systemmodel} we describe ABE-Cities and our adversary models.
In Section~\ref{keyas} we focus on the access policies, whose structure is paramount for the system performance.
In Section~\ref{experimentalevaluation} we evaluate the performance of ABE-Cities.
Section~\ref{conclusions} summarizes our results and concludes the paper.

\section{Related Work}
\label{relatedwork}

Yu et al.~\cite{yu2010achieving} first used ABE techniques for outsourcing sensitive data to a honest-but-curious cloud storage, while enforcing fine-grained access control at the same time.
Their scheme is well applicable for example in ehealth, where patients produce sensitive healthcare records and enjoy full computational capabilities.
However, it is less suitable for IoT applications, in which data is produced by constrained device, which could not have the necessary computational power and energy to produce ABE ciphertexts.
In our system, the IoT devices execute only lightweight operations of symmetric cryptography.
The authors of~\cite{yu2010achieving} also proposed a technique to delegate the burdensome re-encryption of old ciphertexts to the cloud servers, in order to make a revoked key unable to decrypt them.
We use such a re-encryption technique as a building block in our scheme.

Wang et al.~\cite{wang2010hierarchical} proposed a hierarchical ABE scheme that allows the TTP to delegate part of the responsibility to other authorities, which independently make decisions on the semantics of their attributes.
Hierarchical ABE allows also for proxy re-encryption, but it forces the access policies to have a fixed structure, so it limits the flexibility of the access control system.
In this paper, we focused on a non-hierarchical ABE scheme with a single TTP, leaving the possible hierarchical extensions as future work.

Yu et al.~\cite{yu2011fdac} proposed an ABE scheme to encrypt data sensed by a wireless sensor network (WSN).
The authors used broadcast encryption in order to perform key revocations efficiently.
Although their system is suitable for a large set of application scenarios involving locally distributed WSNs, it could be unpractical for geographically distributed ones, like those employed in a smart city.
This is because an actual broadcast is impossible in these scenarios. %
%, and realizing a ``broadcast'' with many unicast transmissions would not scale with the number of sensing devices.
Moreover, the authors propose to perform ABE encryption directly on the sensing devices, which could not have the necessary computational power and energy to do it.
In our system, we propose a key revocation method whose efficiency is %not based on broadcast encryption, but rather on a convenient attribute-based representation of the smart city.
based on  a convenient attribute representation of the smart city.
In addition, in our system the IoT devices execute only lightweight operations of symmetric cryptography.

Yao et al.~\cite{yao2015lightweight} proposed an ABE scheme based on elliptic curves instead of pairings.
This makes ABE operations more lightweight and more suitable for typical IoT constrained devices.
In our scheme, the IoT devices execute only lightweight operations of symmetric cryptography.
This permits us to use the pairing-based ABE scheme of Goyal et al.~\cite{goyal2006attribute}, which allows us to employ advanced features like proxy re-encryption~\cite{yu2010achieving}.

%\textcolor{blue}{
%Recently, Odelu et al.~\cite{odelu2017expressive} proposed an ABE scheme suitable for IoT scenarios because encryption and decryption times are $\mathcal{O}(1)$, and  both ciphertexts and decryption keys are constant in size.
%However, such scheme allows only for access policies formed by AND operators.
%Such poor expressiveness limits the granularity of the access control mechanism, and it might be not sufficient for a smart city scenario.
%On the contrary, our system is based on Goyal's KP-ABE, which allows for any monotonic Boolean formula and offers a greater degree of flexibility and expressiveness.
Recently, Odelu et al.~\cite{odelu2017expressive} proposed an ABE scheme suitable for IoT scenarios because encryption and decryption times are $\mathcal{O}(1)$.
%However, such scheme allows only for access policies formed by AND operators, which limit the granularity of the access control mechanism.
However, such scheme  forces the access policies to have a fixed structure, so it limits the flexibility of the access control system.
On the contrary, our system is based on Goyal's ABE, which offers a greater degree of expressiveness.
%}

\section{Preliminaries}
\label{background}
%
%In this section we give some background on KP-ABE~\cite{goyal2006attribute}, on KP-ABE proxy re-encryption~\cite{yu2010achieving}, and on segment tree structure~\cite{de2000computational} which can be used in ABE to efficiently realize point-in-interval policies~\cite{attrapadung2016attribute}.

\subsection{Key-Policy Attribute-Based Encryption} \label{Sec:KPABE} % [and Access Tree]}
Key-Policy Attribute-Based Encryption (KP-ABE) is a public-key encryption scheme based on bilinear pairings.
It was introduced by Goyal et al.~\cite{goyal2006attribute} as an extension of the original ABE proposed by Sahai and Waters~\cite{sahai2005fuzzy}.
In KP-ABE everyone can encrypt because who encrypts uses only public parameters.
The ciphertexts are labelled with a set of attributes which describes them (\emph{encryption attributes}, $\gamma$).
Decryption keys embed an \emph{access policy} ($\mathcal{T}$), and they are capable of decrypting a ciphertext only if the encryption attributes on the ciphertext match the policy.
Who encrypts data is said a \emph{data producer}, whereas who holds a decryption key and decrypts data is said a \emph{data consumer}.

An access policy is a monotonic Boolean formula on the presence of some attributes on the ciphertext.
It can be represented as a tree in which leaf nodes are attributes (\emph{access policy attributes}, $\lambda$), and non-leaf nodes are AND/OR operators.
%Fig. \ref{fig:ciph-key} shows an example of a ciphertext labelled with some encryption attributes, and a decryption key embedding an access policy.
%\begin{figure}[ht]
%\centering
%\includegraphics[width=.75\columnwidth]{ciph-key.pdf}
%\caption{Example of access policy and encryption attribute set which satisfies the policy in a KP-ABE scheme.}
%\label{fig:ciph-key}
%\end{figure}\\
%The encryption attributes satisfy the policy, so the decryption key is capable of decrypting the ciphertext.
%The access policy in Fig. \ref{fig:ciph-key} reads as follows: ``The attribute A must be present in the ciphertext, and one of the attributes B, C, and D must be present as well.''
For example, let us suppose that a data consumer holds a decryption key embedding the access policy $\mathcal{T} = (A \land (B \lor C \lor D))$ (and thus, its access policy attributes are $\lambda = \{A, B, C, D\}$).
This access policy reads as follows: ``the attribute $A$ must be present in the ciphertext, and one of the attributes $B$, $C$, and $D$ must be present as well.''
A ciphertext labelled, for example, with the encryption attributes $\gamma = \{A, C, E\}$ can be decrypted by the above decryption key.
Goyal's KP-ABE does not allow for non-monotonic access policies, i.e., Boolean formulas including NOT operators.

KP-ABE is resilient to collusion, meaning that two decryption keys cannot be combined somehow to decrypt a ciphertext that they could not decrypt singly.
The set of all the attributes used in a given KP-ABE scheme is the \emph{universe of the attributes} ($\mathcal{U}$) for such scheme.
Without losing in generality, in the following we will indicate an attribute in the universe either with its unique name, or with a unique natural number which is more convenient in formulas.
The attribute sets $\mathcal{U}$, $\gamma$, and $\lambda$ are thus subsets of $\mathbb{N}$.

In order to ease the reading, we abstract away from this paper the mathematical insight of the KP-ABE scheme.
%The quantities $y$, $Y$, and $e^\prime$ will not be discussed because not relevant to this paper.
The interested reader can refer to~\cite{goyal2006attribute} for such details.
We model the KP-ABE scheme with the following black-box primitives.
%\begin{enumerate}
%  \item $(MK, PK) = \mathrm{Setup}(\mathcal{U})$
%  \item $E = \mathrm{Encrypt}(M, \gamma, Y, T_{i \in \gamma})$
%  \item $SK = \mathrm{KeyGen}(MK, \mathcal{T})$
%  \item $M = \mathrm{Decrypt}(E, SK)$
%\end{enumerate}

%\subsubsection{Setup}
%\begin{equation}
%(MK, PK) = \mathrm{Setup}(\mathcal{U}). \nonumber
%\end{equation}
\subsubsection{$(MK, PK) = \mathrm{Setup}(\mathcal{U})$}
This primitive initializes a KP-ABE scheme.
It takes as input the universe of the attributes and generates a random \emph{master key} $MK = (y, t_{i \in \mathcal{U}})$, which is kept secret, and an associated set of \emph{public parameters} $PK = (Y, T_{i \in \mathcal{U}})$.
Each attribute $i$ in the universe is associated to a $t_i$ and a $T_i$.
The $\mathrm{Setup}$ primitive is executed by the TTP.

\subsubsection{$E = \mathrm{Encrypt}(M, \gamma, Y, T_{i \in \gamma})$}
This primitive encrypts a plaintext $M$ with the encryption attributes $\gamma$.
It takes as input $Y$ and $T_{i\in\gamma}$, which are all public parameters. % a set of public parameters
It produces the ciphertext $E = (\gamma, \tilde{e}, e_{i \in \gamma})$.
Each attribute $i$ in the encryption attribute set is associated to a \emph{ciphertext component} $e_i$.
The $\mathrm{Encrypt}$ primitive is executed by a data producer.

\subsubsection{$DK = \mathrm{KeyGen}(MK, \mathcal{T})$}
This primitive generates a new decryption key $DK = (\mathcal{T}, \lambda, dk_{i \in \lambda})$, which is sent to the data consumer in a confidential way.
It takes as input the master key and an access policy $\mathcal{T}$, with access policy attributes $\lambda$.
Each attribute $i$ in the access policy attribute set is associated to a \emph{decryption key component} $dk_i$.
The $\mathrm{KeyGen}$ primitive is executed by the TTP.

\subsubsection{$M = \mathrm{Decrypt}(E, DK)$}
This primitive decrypts the ciphertext $E$ given a decryption key $DK$.
It produces the plaintext $M$ if the decryption key is able to decrypt the ciphertext, $\bot$ otherwise.
The $\mathrm{Decrypt}$ primitive is executed by a data consumer.

\subsection{Proxy Re-Encryption}
Proxy Re-Encryption (PRE) is a technique which allows an entity, given an encrypted message, to produce the same message encrypted with a different key, without accessing the message itself.
Yu et al.~\cite{yu2010achieving} introduced a PRE technique for the KP-ABE scheme.
By using this technique, one can re-encrypt old ciphertexts in order to prevent a revoked key to decrypt them and outsource this burdensome operation to a honest-but-curious CSS without compromising data confidentiality.
%Yu et al.'s PRE is based on \emph{re-encryption keys} that, if applied to ciphertexts, produce new versions of them.
%More in detail, the re-encryption key $rk_{i \leftrightarrow i^\prime}$, which is relative to the attribute $i$, can be applied to the ciphertext component $e_i$ to create a new version $e_i^\prime$ of it.
%The same re-encryption key is also applied to the decryption key components $SK_i$ of all the keys except the revoked one, producing new versions $SK_{i^\prime}$ of them.
%We will refer to the latter procedure as \emph{proxy re-keying}.

Yu et al.'s PRE implements the revocation of a decryption key by means of a system-wide \emph{update} %
%of some of its access policy attributes, in such a way the revoked key cannot decrypt anymore.
of a subset $\mu$ of its access policy attributes ($\mu \subseteq \lambda$).
In particular, the subset $\mu$ is such that, without those attributes in a ciphertext, the access policy will surely not be satisfied. %the minimal set of attributes (PER YU).
Updating an attribute $i$ to a new \emph{version} simply means that all the relative cryptographic quantities in the system are changed.
Specifically, for all the attributes $i \in \mu$, the quantities $t_i$, $T_i$, the $e_i$'s of all the old ciphertexts, and the $dk_i$'s of all the decryption keys except those of the revoked key are updated with new quantities $t_i^\prime$, $T_i^\prime$, $e_i^\prime$, and $dk_i^\prime$.
%The new $T_i^\prime$ quantity must be used from now on to produce new ciphertexts labelled with the encryption attribute $i$.
%The new $t_i^\prime$ quantity will be used from now on by the TTP to produce new decryption keys that have the decryption attribute i in their access policy.
The burdensome computation of the updated $dk_i^\prime$ can be outsourced to the CSS as well.
From now on, we implicitly consider the quantities $t_i$, $T_i$, $e_i$, and $dk_i$ always accompanied by the information about their versions.

The computation of the $e_i^\prime$'s and the $dk_i^\prime$'s is delegated to a honest-but-curious CSS.
To do this, the TTP computes a \emph{re-encryption key} ($rk_i$) and sends it to the CSS, which can apply it to an $e_i$ to obtain an $e_i^\prime$, and to a $dk_{i}$ to obtain a $dk_i^\prime$.
The computation of the $dk_i^\prime$'s cannot be completely outsourced, since the CSS would then know all the decryption keys, so it could violate the data confidentiality.
To avoid this, the CSS is provided with all the decryption key components except those relative to a special attribute (\emph{dummy attribute}) which has no meaning and is never updated to a new version.
The dummy attribute is ANDed at the root of the access policies of all the decryption keys, and it is included as an encryption attribute in all the ciphertexts.
In this way, a decryption key cannot decrypt anything without the component relative to the dummy attribute, which is known only by the data consumer.
Thus, only the data consumer can effectively use his/her decryption key.

Proxy re-encryption is performed in a lazy fashion (\emph{lazy re-encryption}), meaning that the CSS does not perform any burdensome operation at the moment of the key revocation, but only afterwards when needed.
At the moment of revocation, the TTP produces a re-encryption key for each attribute $i$ in $\mu$ and adds it to a \emph{re-encryption key history} ($RKH_i$) stored in the CSS.
$RKH_i$ is a list that keeps track of all the re-encryption keys relative to the attribute $i$, one for each version update of the attribute ($rk_i, rk_i^\prime, rk_i^{\prime\prime}, \dots$).
%The first version of every attribute is zero, thus the number of entries in the re-encryption key history gives the latest version of the attribute.
%When a user asks the CSP to retrieve a certain ciphertext whose components are outdated, the CSP applies proxy re-encryption on them.
%When a user asks the CSP for a ciphertext whose components are outdated, the CSP applies proxy re-encryption on them.
%If the attribute $i$ has been updated multiple times in the meanwhile, the CSP will re-encrypt the corresponding component using multiple re-encryption keys, all of which are stored in the attribute history list.
%Moreover, if the user's decryption key have outdated components, the CSP applies proxy re-keying on them.
Only when a data consumer asks the CSS for a ciphertext whose components are outdated, then the CSS computes the new components.
If a ciphertext component was outdated of more than one version, then the CSS will use more than one re-encryption key from the history to update it.
Similarly, if the consumer's decryption key has outdated components, then the CSS computes the new components.
If a decryption key component was outdated of more than one version, then the CSS will use more than one re-encryption key from the history to update it.
Finally, the CSS provides the data consumer with the ciphertext and, possibly, with the new decryption key components.
Note that the new decryption key components can be transmitted to the data consumer in the clear, without compromising the secrecy of the key.
Indeed, an adversary cannot use the decryption key components unless she knows also the one relative to the dummy attribute, which is never updated.

In order to ease the reading, we abstract away from this paper the mathematical insight of the PRE scheme.
The interested reader can refer to~\cite{yu2010achieving} for such details.
We model the PRE scheme with the following black-box primitives.

\subsubsection{$(t_i^\prime, T_i^\prime, rk_i) = \mathrm{UpdateAttribute}(i, MK)$}
This primitive updates the attribute $i$, meaning that it produces the new quantities $t_i^\prime$, $T_i^\prime$, and the re-encryption key $rk_i$.
The $\mathrm{UpdateAttribute}$ primitive is executed by the TTP.

\subsubsection{$e_i^\prime = \mathrm{UpdateE}(i,e_i,RKH_i)$}
This primitive updates a ciphertext component to a new version.
%It takes as input the attribute $i$, the ciphertext component $e_i$, and the list $AHL_i$ which contains the re-encryption key $rk_{i \leftrightarrow i^\prime}$.
%Such re-encryption key is combined with the ciphertext component $e_i$ to output $e_i^\prime$.
The $\mathrm{UpdateE}$ primitive is executed by the CSS.

\subsubsection{$dk_i^\prime = \mathrm{UpdateDK}(i, dk_i, RKH_i)$}
This primitive updates a decryption key component to a new version.
%It takes as input the attribute $i$, the decryption key component $SK_i$, and the list $AHL_i$ which contains the re-encryption key $rk_{i \leftrightarrow i^\prime}$.
%Such re-encryption key is combined with the decryption key component $SK_i$ to output $SK_{i^\prime}$.
The $\mathrm{UpdateDK}$ primitive is executed by the CSS.

\subsection{Segment Trees}

A segment tree is a data structure that represents a set of intervals~\cite{de2000computational}.
It allows us to efficiently query which intervals contain a given point.
Segment trees are useful in ABE schemes to implement efficient point-in-interval access policies~\cite{attrapadung2016attribute}.
Although segment trees can in principle represent any point and interval in $\mathbb{R}$, in this paper we focus on \emph{discrete} segment trees, i.e., segment trees that represent points and intervals in $\mathbb{Z}$.
Specifically, we focus on discrete segment trees capable of representing any point and any interval included in a limited number range $[1,\rho]$, with $\rho \in \mathbb{N}$.
In the following, we will generically use the term ``segment tree'' to intend such a type of discrete segment tree.

\begin{figure}[t]
\centering
\includegraphics[width=.6\columnwidth]{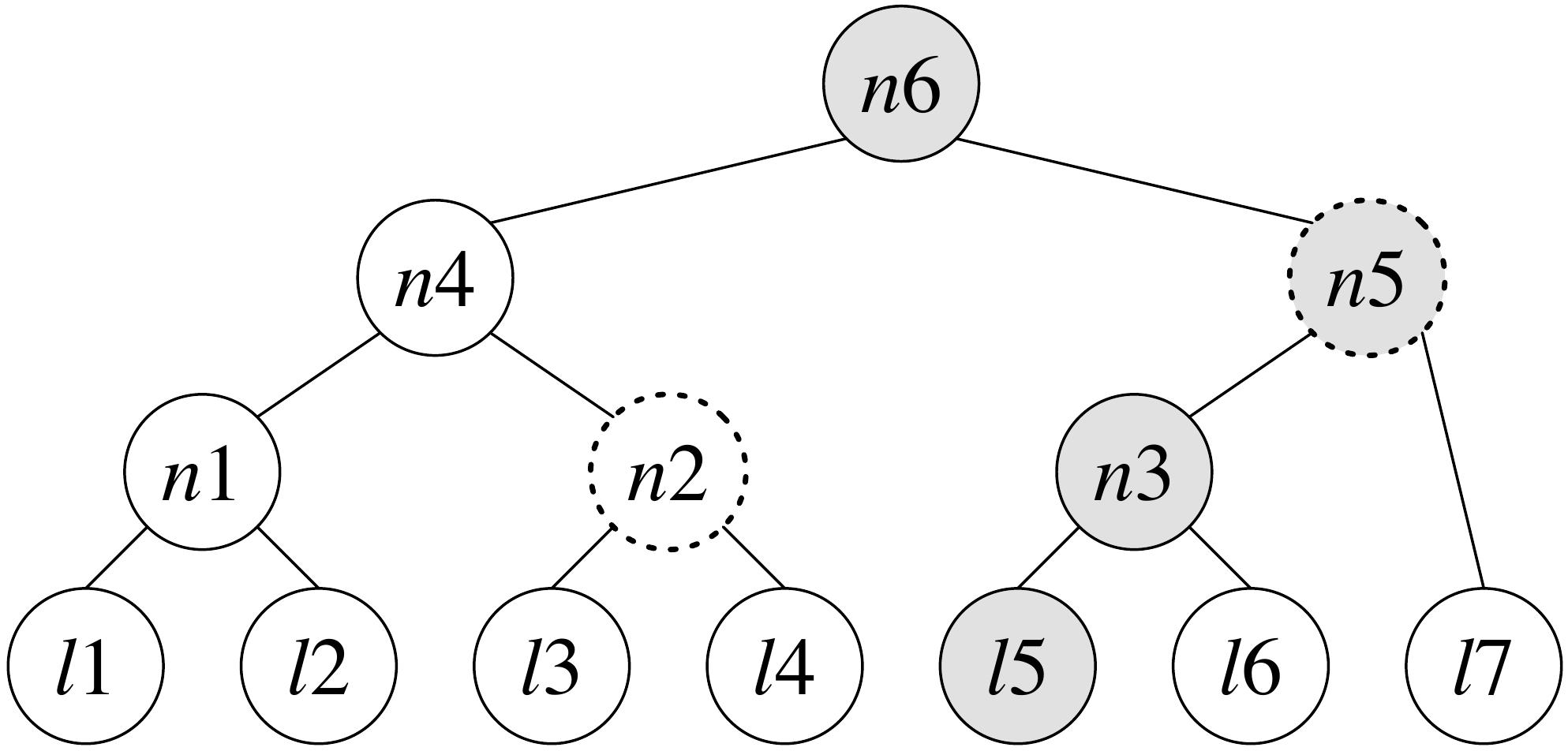}
\caption{
Example of a segment tree.
Grey nodes form the point representation set of $5$, dashed-border nodes form the interval representation set of $[3,7]$.
}
\label{fig:segtree}
\end{figure}

%A segment tree is a binary tree in which the leaf nodes are all the integer numbers within a given range $[1,\rho]$.
A segment tree over the range $[1,\rho]$ is a binary tree in which each number in $[1,\rho]$ is relative to a leaf.
For example, Fig. \ref{fig:segtree} shows a segment tree over the range $[1,7]$.
The leaf $l1$ is relative to the number $1$, the leaf $l2$ to the number $2$, and so on.
%The representation set for the value $\nu=5$ contains the nodes set $RS_5=\{5,10,12,13\}$, i.e., all the nodes in the path from the leaf node $\nu$ to the root.
%The representation set for the interval $\iota=[3,7]$ contains the nodes set $RS_{[3,7]}=\{9,12\}$, i.e., the minimum set of nodes whose subtrees include all-and-only the values of the interval $\iota$.
%The intersection $RS_\nu \cap RS_\iota$ is not an empty set. Hence, we conclude that the value $\nu$ is contained in the interval $\iota$.
A generic point $\nu$ in the range $[1,\rho]$ is represented by a \emph{point representation set} $RS_\nu$, which is formed by the nodes on the path from the root to the leaf relative to the point.
For example, in the segment tree of Fig. \ref{fig:segtree}, the point $\nu=5$ is represented by the point representation set $RS_\nu=\{n6, n5, n3, l5\}$.
It can be shown that every point representation set is $\mathcal{O}(\log \rho)$ nodes.
On the other hand, a generic interval $\iota$ included in the range is represented by an \emph{interval representation set} $RS_\iota$, which is the minimum set of nodes whose descendant leaves form the interval.
For example, in the segment tree of Fig. \ref{fig:segtree}, the interval $\iota=[3,7]$ is represented by the interval representation set $RS_\iota=\{n2, n5\}$.
It can be shown that every interval representation set is $\mathcal{O}(\log \rho)$ nodes.
By construction, $RS_\nu$ and $RS_\iota$ have non-empty intersection iff the point $\nu$ belongs to the interval $\iota$: $\nu \in \iota \Leftrightarrow RS_\nu \cap RS_\iota \neq \emptyset$.
For example, in the segment tree of Fig. \ref{fig:segtree}, the point $\nu=5$ belongs to the interval $\iota=[3,7]$ because $RS_\nu \cap RS_\iota = \{n5\} \neq \emptyset$.

Segment trees are useful in ABE schemes to implement efficient point-in-interval access policies~\cite{attrapadung2016attribute}, i.e., access policies which are satisfied iff a given point $\nu$ belongs to a given interval $\iota$.
According to the terminology in~\cite{attrapadung2016attribute}, we consider only ``KP-ABE Type 1 construction'', which means that the ciphertext embeds the point, and the access policy embeds the interval.
The point-in-interval access policy is thus implemented in the following way.
Each node of the segment tree is represented by an ABE attribute.
The ciphertext is labelled with the attributes representing the nodes of the point representation set $RS_\nu$.
The access policy is an OR operator between the nodes representing the interval representation set $RS_\iota$.
By construction, the access policy is satisfied iff $RS_\nu$ and $RS_\iota$ have non-empty intersection, that is, iff $\nu$ is in $\iota$.

%In a basic representation, a value to be checked is represented by a single attribute, and the key access structure contains as many attributes as the number of values in an interval to represent, so the number of attributes of the key (\emph{key size}) increases linearly.
%On the contrary, the efficiency of the implementation based on segment tree is that it introduces only logarithmic overhead on the key size.
%The number of attributes used to label a ciphertext increases logarithmically as well.

\section{Proposed Scheme}\label{systemmodel}

In our system, a city is represented by a \emph{street network}, i.e., a graph in which the edges represent \emph{streets}.
Each street is characterized by an identifier and is partitioned into \emph{road segments}.
A \emph{sensing device} is a possibly constrained device placed on a road segment that acquires data (\emph{sensed data}, $SD$) relative to such a road segment.
The sensed data is stored in an encrypted form in a \emph{Cloud Storage Service} (CSS), where it is made read-only accessible to the \emph{users}.
The users are the data consumers of our system.

A user can be authorized to access only data sensed from some regions of the city, specified as a set of road segments.
For example, supposing the sensing devices to be cameras, the user can monitor the traffic congestion on the route he/she usually travels from home to work.
To do this, the cameras could store in the CSS many short video files to implement an encrypted streaming service.
%Alternatively, if videos are considered too privacy-sensitive, cameras could process them to extract the traffic average speed on that road segment and store only this information in the CSS.
The user can also monitor multiple routes, in order to identify the least congested one day by day.
Moreover, some advanced-privilege users could be provided, for example a transportation management service which can monitor all the road segments of the city.

\begin{figure}[ht]
\includegraphics[width=\columnwidth]{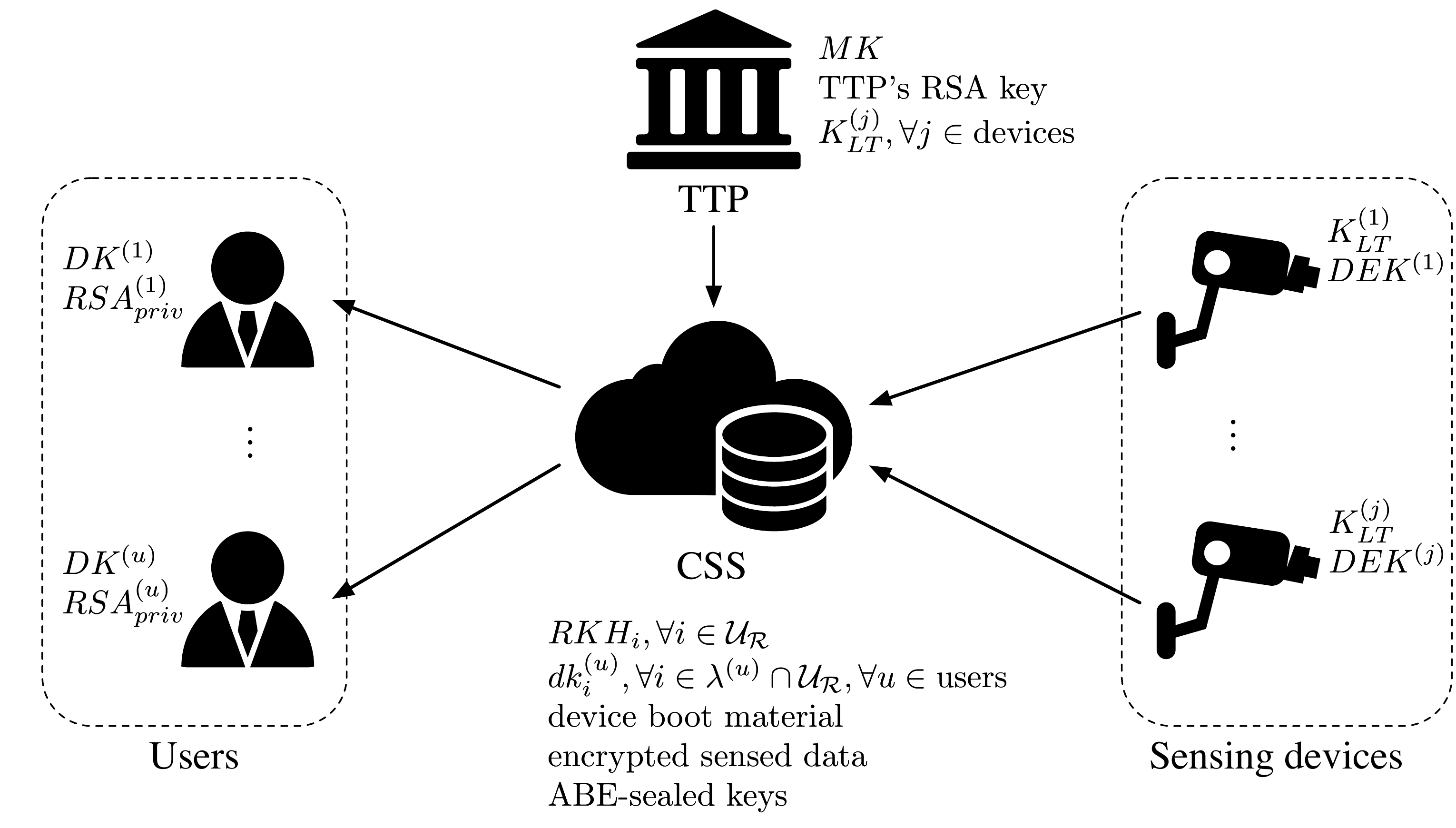}
\caption{System architecture.}
\label{fig:architecture}
\end{figure}

The architecture of our system is shown in Fig. \ref{fig:architecture}.
\\
In the following, we will give a general and intuitive description of the system while in Section \ref{ssec:procedures} we will describe the system procedures in detail.
The TTP holds the KP-ABE master key and an RSA key which it uses to sign messages.
There is nothing special in choosing RSA as digital signature algorithm, elliptic-curve digital signature algorithms (EC-DSA) are suitable as well.
The notation $\operatorname{Sign}(\cdot)$ is used to represent the TTP's signature on something.
The TTP also shares with each sensing device $j$ a \emph{long-term symmetric key} ($K_{LT}^{(j)}$), which is supposed to be preloaded in the sensing devices.
Each sensing device holds also a \emph{data encryption key} ($DEK^{(j)}$), which is the actual symmetric key used to encrypt the sensed data.
The data encryption key is frequently renewed by the sensing device, once for each produced piece of data.
Each user $u$ holds a KP-ABE decryption key $DK^{(u)}$, and an RSA key $RSA_{priv}^{(u)}$ which he/she uses to decrypt messages received from the TTP.
We denote by $RSA_{pub}^{(u)}$ the corresponding public key.
The CSS maintains a database of re-encryption key histories and decryption key components to perform proxy re-encryption.
In addition, it stores \emph{device boot material}, \emph{encrypted sensed data}, and \emph{ABE-sealed keys}.
The device boot material is needed by the sensing devices to encrypt their data.
The encrypted sensed data is the actual data produced by the sensing devices and encrypted with a symmetric encryption algorithm.
The ABE-sealed keys are necessary to decrypt the sensed data and are in turn encrypted (\emph{sealed}) with KP-ABE.
The CSS stores an ABE-sealed key for each sensing device and for each day of system operation.
To access a piece of sensed data, users must first decrypt the corresponding ABE-sealed key and then decrypt the sensed data.

%The sensing devices execute only symmetric cryptography operations.
%KP-ABE encryption, which is computationally burdensome, is delegated to the TTP.
%Therefore, our system can employ also resource-constrained sensing devices, which makes it suitable for a broader set of IoT smart city scenarios.

\begin{figure}[t]
\centering
\includegraphics[width=\columnwidth]{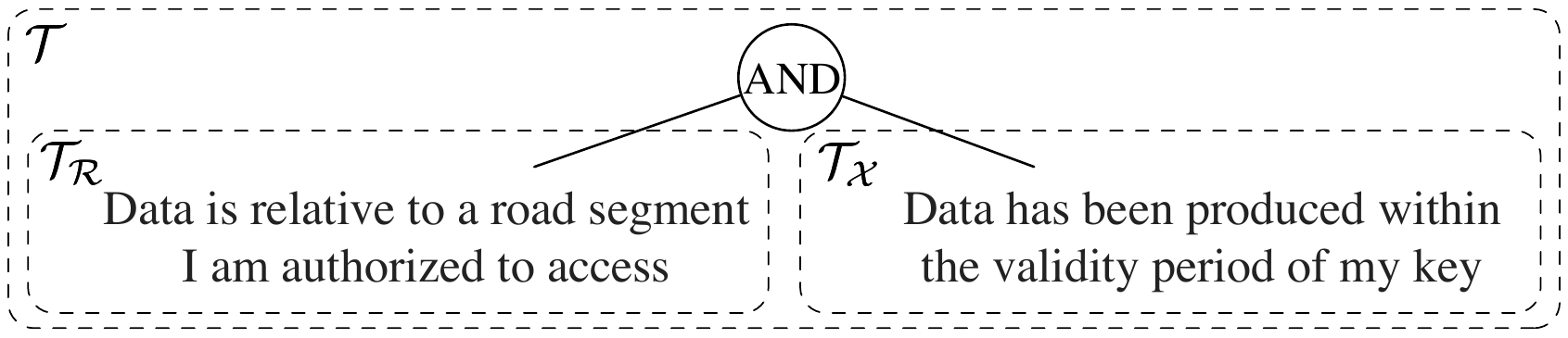}
\caption{Example of access policy for a user's decryption key in our system. In order to decrypt an ABE-sealed key, the user's decryption key must satisfy both the subtrees.}
\label{fig:key}
\end{figure}

The universe of the attributes $\mathcal{U}$ is logically divided into two subsets: $\mathcal{U}_\mathcal{R}$ and $\mathcal{U}_\mathcal{X}$.
$\mathcal{U}_\mathcal{R}$ includes the attributes to represent road segments.
$\mathcal{U}_\mathcal{X}$ includes the attributes to represent time.
Each ABE-sealed key relative to the sensing device $j$ is labelled with the encryption attributes $\gamma^{(j)}$, which are logically divided into two subsets: $\gamma_\mathcal{R}^{(j)}$ and $\gamma_\mathcal{X}^{(j)}$.
The attributes in $\gamma_\mathcal{R}^{(j)}$ identify the device's road segment, while the attributes in $\gamma_\mathcal{X}^{(j)}$ identify the day in which data has been sensed (\emph{data production date}).
Similarly, the access policy $\mathcal{T}$ embedded in each decryption key is logically divided into two subtrees: $\mathcal{T}_\mathcal{R}$ and $\mathcal{T}_\mathcal{X}$ (Fig. \ref{fig:key}).
The two subtrees are operands of an AND operator ($\mathcal{T} = \mathcal{T}_\mathcal{R} \land \mathcal{T}_\mathcal{X}$), which is the root of the access policy.
The $\mathcal{T}_\mathcal{R}$ subtree identifies the road segments the user is authorized to monitor (\emph{authorized road segments}) while $\mathcal{T}_\mathcal{X}$ specifies the \emph{validity period} of the key.
The validity period is the period of time from the day the user joins the system to the day in which his/her subscription ends.
%The access policy determines which road segments the user is authorized to monitor (authorized road segments) and specifies the validity period of the decryption key.
Hence, the user is able to decrypt sensed data produced during his/her validity period by sensing devices belonging to his/her authorized road segments.

Note that, in order to implement proxy re-encryption, the CSS is let know of
%TTP outsources to the CSS 
all the re-encryption keys except those relative to the attributes in $\mathcal{U}_\mathcal{X}$, and 
all the decryption key components except those relative to the $\mathcal{T}_\mathcal{X}$ subtree.
The presence of at least one attribute in $\mathcal{U}_\mathcal{X}$ is ANDed at the root of the access policy of all the decryption keys.
%In addition, the CSS is let know of all the re-encryption keys except those relative to the attributes in $\mathcal{U}_\mathcal{X}$.
In practice, the attributes in $\mathcal{U}_\mathcal{X}$, which are never updated to a new version, replace the dummy attribute of the Yu et al.'s PRE scheme~\cite{yu2010achieving}.

%Note that the TTP is not required of high bandwidth or high availability.
%The effort in data exchanging is totally delegated to the CSS, which receives encrypted sensed data from the sensing devices and transmits it to the users.

%The access policy of each decryption key establishes what data the user is authorized to decrypt.

%The access policy set contains attributes from both the subsets $\mathcal{U}_\mathcal{R}$ and $\mathcal{U}_\mathcal{X}$.
%The access policy

%The decryption key is capable of decrypting data from a subset of road segments produced during the \emph{validity period} of the key. %, and thus that one of the subscription.
%Indeed, each decryption key corresponds to a subscription, so its validity has to be limited to a given time period, which is specified in the access policy.

%%%%We logically divide the universe of the attributes $\mathcal{U}$ into two subsets.
%%%%One is related to the representation of the street network, and we call it $\mathcal{U}_\mathcal{R}$.
%%%%The other subset, which we call $\mathcal{U}_\mathcal{X}$, is related to the representation of the time.
%Specifically, the attributes in $\mathcal{U}_\mathcal{X}$ are used both to label the ABE ciphertexts with the data production date and to define the validity period of a decryption key inside its access policy.
%Each ABE ciphertext is labelled with attributes from both the subsets $\mathcal{U}_\mathcal{R}$ and $\mathcal{U}_\mathcal{X}$.

\subsection{Adversary Model(s) and Security Analysis}
We consider three types of adversary: (i) the honest-but-curious CSS, (ii) an external adversary capable of eavesdropping traffic and compromising sensing devices, and (iii) a set of colluding users wanting to gain more authorizations illegally.
The CSS is assumed to be honest-but-curious,
%meaning that it faithfully respects the specifications, but it is also interested in reading the stored sensed data.
meaning that it trustworthy carries out data distribution and elaboration functions but is interested in reading the stored sensed data.
%With respect to this adversary model, our system offers similar security properties of Yu et al. \cite{yu2010achieving}.
To decrypt the stored sensed data, the CSS should first decrypt the ABE-sealed key, which is encrypted with KP-ABE.
However, the CSS does not know any \emph{complete} decryption key to do this.
In particular, it knows none of the decryption key components relative to the attributes in $\mathcal{U}_\mathcal{X}$, whose presence is always ANDed at the root of all the access policies.

The external adversary can eavesdrop all the traffic between the entities of our system.
Her goal is to read the sensed data.
To achieve this, she could either steal decryption keys, or inject malicious device boot material to make the sensing devices encrypt data with a compromised key.
However, none of these tactics are viable.
Stealing decryption keys is infeasible since the TTP sends them to newly joined users encrypted with RSA.
Injecting malicious device boot material to a sensing device is infeasible too, since it is authenticated with the long-term secret key shared between the device and the TTP.
%Moreover, the CSS itself could prevent this as it
Even if the external adversary compromises a sensing device, she can only read the new sensed data produced by such device after the compromise, but not the old one.
Indeed, as we will explain in detail in Section \ref{ssec:procedures}, each device changes the data encryption key at each new piece of produced data.
At the moment of compromise, only the current data encryption key is actually compromised.
The old data encryption keys are securely destroyed by the sensing device and are not recoverable from the current one.

The third type of adversary is a set of users who collude to gain more authorizations illegally.
This is infeasible because KP-ABE is resilient to collusion, meaning that two decryption keys cannot be combined somehow to decrypt an ABE-sealed key that they could not decrypt singly.

\subsection{System Procedures} \label{sec:procedures}
\label{ssec:procedures}
%%%%%%%%%%%%%%                      SETUP PROCEDURE                         %%%%%%%%%%%%%%
\subsubsection{Setup Procedure}
This procedure initializes the system.

The TTP defines the universe of the attributes $\mathcal{U} = \mathcal{U}_\mathcal{R} \cup \mathcal{U}_\mathcal{X}$ and then executes the $\mathrm{Setup}(\mathcal{U})$ primitive which produces the master key and the public parameters.
The TTP keeps the master key secret and stores in the CSS an empty re-encryption key history for each attribute in $\mathcal{U}_\mathcal{R}$.

%%%%%%%%%%%%%%             KEY DISTRIBUTION PROCEDURE               %%%%%%%%%%%%%%
\subsubsection{Key Distribution Procedure}
This procedure provides a user with a new decryption key.
It is executed when a new user joins the system, as well as when an old user needs to renew his/her decryption key because it expired or it has been compromised.

The TTP first defines the access policy of the new user $u$.
Which access policy assigning to each user depends strictly on the application.
Once the TTP defined the user's access policy $\mathcal{T}$, it creates the user's decryption key $DK^{(u)}$ by executing the $\mathrm{KeyGen}(MK, \mathcal{T})$ primitive.
The TTP encrypts the message $(DK^{(u)}, \text{cur.date}, \operatorname{Sign}(DK^{(u)}, \text{cur.date}))$, where cur.date is the current date, with the user's public key and provides it to the user either with a direct channel or through the CSS.
The user decrypts it and verifies the TTP signature to be valid, and cur.date to be the actual current date.
If everything is correct, the user accepts $DK^{(u)}$ as his/her decryption key.
In the meanwhile, the TTP sends to the CSS the message $(u, dk_{i\in\lambda^{(u)} \cap \mathcal{U}_\mathcal{R}}, \text{cur.date}, \operatorname{Sign}(u, dk_{i\in\lambda^{(u)} \cap \mathcal{U}_\mathcal{R}}, \text{cur.date}))$, where $\lambda^{(u)}$ are the access policy attributes of the user.
The CSS verifies the TTP signature to be valid, and cur.date to be the actual current date.
If everything is correct, the CSS stores the decryption key components $dk_{i\in\lambda^{(u)} \cap \mathcal{U}_\mathcal{R}}$. %and associates them to the user $u$.

%%%%%%%%%%%%%%%%%              SEAL PROCEDURE               %%%%%%%%%%%%%%%%
\subsubsection{Seal Procedure}
This procedure is executed once a day at the midnight. %or after a key revocation

For each sensing device $j$, the TTP generates a random data encryption key $DEK^{(j)}$ and encrypts it with the $\mathrm{Encrypt}(DEK^{(j)}, \gamma^{(j)}, PK)$ primitive.
The result constitutes the ABE-sealed key ($ASK^{(j)}$) for the sensing device $j$, and it is stored in the CSS.
The CSS stores an ABE-sealed key for each sensing device and for each day of system operation.
Also, for each sensing device $j$ the TTP encrypts with the long-term symmetric key $K_{LT}^{(j)}$ the message $(DEK^{(j)}, \text{cur.date}, \operatorname{MAC}_{K_{LT}^{(j)}}(DEK^{(j)}, \text{cur.date}))$, where $\operatorname{MAC}_{K_{LT}^{(j)}}(\cdot)$ denotes a message authentication code keyed by $K_{LT}^{(j)}$.
The result constitutes the device boot material for the sensing device $j$, and it is stored in the CSS too.
The CSS stores a device boot material for each sensing device.
Each sensing device retrieves its device boot material and checks the $\operatorname{MAC}$ to be valid, and cur.date to be the actual current date.
If everything is correct, the device accepts $DEK^{(j)}$ as its current data encryption key and initializes a \emph{data encryption counter} ($c^{(j)}$) to zero.

%%%%%%%%%%%%%%             DATA PRODUCTION PROCEDURE               %%%%%%%%%%%%%%
\subsubsection{Data Production Procedure}

This procedure is executed every time a sensing device produces a new piece of data.

First, the device encrypts the sensed data $SD^{(j)}$ with its current data encryption key $DEK^{(j)}$.
The result constitutes an encrypted sensed data $ESD^{(j)}$, which is stored in the CSS together with the current data encryption counter $c^{(j)}$.
The CSS stores many tuples $(ESD^{(j)}, c^{(j)})$ for each sensing device.
Then, the sensing device computes a new data encryption key as a one-way hash of the old one: $DEK^{(j)} \leftarrow \operatorname{H}(DEK^{(j)})$.
Finally, the device securely destroys the old data encryption key and increments its data encryption counter.
This key renewal method prevents an external adversary who physically compromises a sensing device from decrypting old encrypted sensed data, thus ensuring backward secrecy.

%%%%%%%%%%%%%%             DATA CONSUMPTION PROCEDURE               %%%%%%%%%%%%%%
\subsubsection{Data Consumption Procedure} \label{sec:dataConsProc}

This procedure is executed every time a user wants to access a piece of sensed data.
The user $u$ first sends a data request to the CSS specifying which sensed data he/she wants to access.
On receiving the data request, the CSS checks whether some of the stored decryption key components are out of date by comparing their versions with the latest version of the relative attribute.
For each out-of-date component $dk_i$, the CSS updates it by executing: $dk_i^\prime\,=\,\mathrm{UpdateDK}(i, dk_i, RKH_i)$, and it provides $dk_i^\prime$ to the user.
Then, the CSS checks whether some of the ciphertext components of the ABE-sealed key $ASK^{(j)}$ relative to the sensed data requested by the user are out of date.
For each out-of-date component $e_i$, the CSS updates it by executing: $e_i^\prime\,=\,\mathrm{UpdateE}(i, e_i, RKH_i)$, and it replaces $e_i$ with the updated version.
Finally, the CSS provides the ABE-sealed key $ASK^{(j)}$ and the tuple $(ESD^{(j)}, c^{(j)})$ relative to the requested data to the user.
On receiving this, the user decrypts the ABE-sealed key by executing the $\mathrm{Decrypt}(ASK^{(j)}, DK^{(u)})$ primitive.
Of course, if the user is not authorized to access the requested data, such primitive will return $\bot$.
Otherwise, the user applies $c^{(j)}$ times the one-way hash $\operatorname{H}(\cdot)$ on the result, thus obtaining the data encryption key $DEK^{(j)}$ with which the encrypted sensed data $ESD^{(j)}$ was encrypted.
The user can thus read the sensed data $SD$.

%%%%%%%%%%%%%%                KEY REVOCATION PROCEDURE                %%%%%%%%%%%%%%
\subsubsection{Key Revocation Procedure}

Whenever a user $u$'s decryption key must be revoked, for example when his/her key is compromised,
%authorization changes or his/her decryption key is compromised, 
the system makes it ineffective to decrypt any ABE-sealed key by executing the key revocation procedure.
%must revoke the related decryption key, which means to make such decryption key useless.

At first, the TTP determines the subset $\mu^{(u)} \subseteq \lambda^{(u)}$, i.e., a set of attributes without which the access policy will never be satisfied.
%which contains the attributes to update.
% set of attributes without which the access policy will never be satisfied.
The TTP must update such attributes in order to revoke the user $u$.
In our system, the subset $\mu^{(u)}$ is formed by the attributes $\lambda^{(u)} \cap \mathcal{U}_\mathcal{R}$.
For each attribute $i\in \mu^{(u)}$, the TTP updates the related quantities by executing: $(t_i^\prime, T_i^\prime,  rk_i) = \mathrm{UpdateAttribute}(i, MK)$, and it replaces $t_i$ and $T_i$ with the updated versions.
Then, the TTP sends the message $(u, rk_{i, \forall i\in \mu^{(u)}}, \text{cur.date}, \operatorname{Sign}(u, rk_{i, \forall i\in \mu^{(u)}}, \text{cur.date}))$ to the CSS.
The CSS verifies the TTP signature to be valid, and cur.date to be the actual current date. 
If everything is correct, the CSS adds each re-encryption key $rk_i$ to the proper $RKH_i$ 
and erases all the decryption key components $dk_{i\in\lambda^{(u)} \cap \mathcal{U}_\mathcal{R}}$ related to the revoked user $u$. %from the memory
Finally, the TTP executes a seal procedure.
The user is now revoked, and his/her key is not capable of decrypting any ABE-sealed key anymore.

\section{Universe of the Attributes and Access Policies}\label{keyas}

\subsection{Road Segments Representation: Universe Subset $\mathcal{U}_\mathcal{R}$}

\subsubsection{Basic Representation}

The simplest representation consists in mapping each road segment onto one attribute.
In this way, the $\gamma_\mathcal{R}^{(j)}$ set of each ABE-sealed key is formed by just one attribute.
On the other hand, the $\mathcal{T}_\mathcal{R}$ subtree of a decryption key is an OR between many attributes, one for each road segment the user is authorized to monitor.
We will refer to this implementation as the \emph{basic representation}.

By using this representation, the number of attributes of the key access policy, i.e., its \emph{key size}, grows linearly with the number of authorized road segments.
When a user is revoked, the attributes in the $\mathcal{T}_\mathcal{R}$ subtree of his/her key access policy must be updated.
The users whose decryption keys shared any attribute, and thus any authorized road segment,
%-- in the $\mathcal{T}_\mathcal{R}$ subtree --
with the revoked key are called \emph{affected users}.
An affected user is involved in the revocation process of another user, and his/her key must be updated by the CSS during the data consumption procedure, as explained in Section \ref{sec:procedures}.
Our aim is making the key revocation more efficient by limiting the number of affected users.
This improvement lightens the load on the CSS in terms of proxy re-encryption operations.
%We want to limit the number of affected users to lighten the load on the CSS, but also to speed up the data consumption procedure.

\subsubsection{Segment Tree Representation}

In the basic representation, it is very likely that two users share some road segments, and, in this case, if one of them is revoked, the other will be surely an affected user, thus generating a \emph{collision}. %, and his/her decryption key must be updated.
For example, a user authorized to monitor an entire street will collide with any other user authorized to monitor any subset of road segments of that street.

Segment trees, already used in ABE to reduce the key size~\cite{attrapadung2016attribute}, can help us to reduce the average number of affected users, too.
For example, a user authorized to monitor an entire street will not collide with another user authorized to monitor only some road segments of that street.
%Since each street is partitioned into road segments,
%and a user is likely to subscribe to consecutive road segments possibly belonging to the same street, we decide to represent each street with a segment tree.
%We create a segment tree for each way in the system.
Let $\rho$ be the number of road segments of a generic street.
We denote road segments of that street with identifiers from $1$ to $\rho$ and build a segment tree in which leaf nodes are the road segment identifiers.
Each node of the segment tree is represented by an ABE attribute.
%Note that all the $2\rho-1$ nodes of the segment tree are attributes.

\begin{figure}[ht]
\centering
\includegraphics[width=.7\columnwidth]{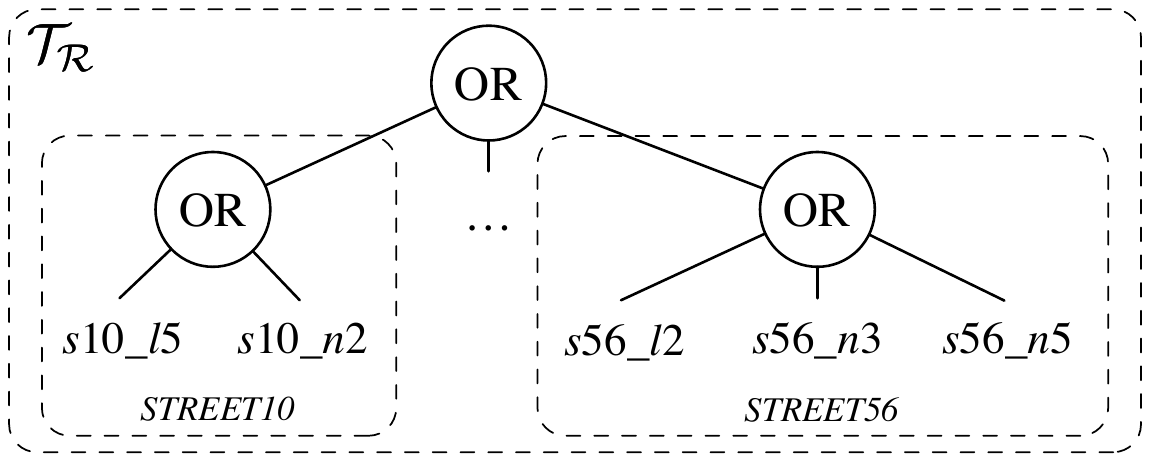}
\caption{Example of $\mathcal{T}_\mathcal{R}$ subtree of the key access policy with segment tree representation.
%It can be represented by using a single OR operator.
The subtree $s10\_l5 \lor s10\_n2$  is a point-in-interval access policy of consecutive road segments of the street \textit{STREET10}.}
\label{fig:R_segtree}
\end{figure}

%Each ABE-sealed key is labelled with point representation set of $\mathcal{O}(\log \rho)$ attributes concerning the attribute set $\gamma_\mathcal{R}^{(j)}$.
The $\gamma_\mathcal{R}^{(j)}$ set of each ABE-sealed key is formed by a point representation set and contains $\mathcal{O}(\log \rho)$ attributes.
%If a user is authorized to monitor some adjacent road segments of a street, an interval representation set is created.
Thus, the TTP is required of a little more effort to produce the ABE-sealed keys since the complexity of the $\mathrm{Encrypt}$ primitive depends on the size of the encryption attribute set $\gamma$.
On the other hand, we use a point-in-interval access policy to identify the subset of consecutive road segments of a street which a user is authorized to monitor.
The $\mathcal{T}_\mathcal{R}$ subtree of an access policy is an OR between many point-in-interval access policies, one for each street the user is authorized to monitor.
Fig. \ref{fig:R_segtree} shows a graphic example.
%The $\mathcal{T}_\mathcal{R}$ subtree of an access policy consists of as many interval representation sets as the different streets the user is authorized to monitor, and Fig. \ref{fig:R_segtree} shows a graphic example.
\\
We will refer to this implementation as the \emph{segment tree representation}.

In addition to reduce the average number of affected users, this representation reduces the key size if compared to the basic representation.
Indeed, the size of an interval representation set is always less than or equal to the number of represented elements.
%Therefore, this representation reduces the key size if compared to the basic representation.
%Moreover, collisions among users who share the same road segment are not certain to happen.
%Collisions can occur when two users share the same attribute of an interval representation set, and we expect that this happens less frequently.
%For example, a user authorized to monitor an entire street will not collide with another user authorized to monitor a part of that street. %Rivedere.
%On the other hand, each ABE-sealed key is labelled with point representation set of $\mathcal{O}(\log \rho)$ attributes concerning the attribute set $\gamma_\mathcal{R}^{(j)}$.

\subsubsection{Attribute-Pool Representation}
%Pericle, ho cercato di farla più formale però si capisce anche meno...
% Appena sotto c'è la vecchia versione tutta commentata.
We introduce a third representation which extends the segment tree representation to further reduce the number of affected users due to a key revocation.
We replace each attribute $i$ in the universe subset $\mathcal{U}_\mathcal{R}$ of the segment tree representation with a pool of $\varepsilon$ attributes $\{i_\omega\}$, with $\omega \in [1,\varepsilon]$ (\emph{replicas}).
%For each attribute of the universe subset $\mathcal{U}_\mathcal{R}$ in the segment tree representation, we define $\varepsilon$ equivalent copies, in the new representation.
%In other words, we replace each attribute $i$ in $\mathcal{U}_\mathcal{R}$ with a pool of $\varepsilon$ attributes
%$\Omega_i = \{i_\omega\}_{\omega=1}^\varepsilon$. 
Each replica $i_\omega$ in the pool has the same meaning. % it is simply an alternative.
%$i_1,i_2,\dots, i_\epsilon$, each of which has the same semantic value.rk_i
%So, the universe subset $\mathcal{U}_\mathcal{R}$ grows by a factor $\epsilon$ if compared to the one of the segment tree representation.

%To identify a generic road segment,
The $\gamma_\mathcal{R}^{(j)}$ set of each ABE-sealed key is formed by $\varepsilon$ point representation sets and contains $\mathcal{O}(\varepsilon \log \rho)$ attributes.
%Thus, the TTP is required of a little more effort to produce the ABE-sealed keys since the complexity of the $\mathrm{Encrypt}$ primitive depends on the size of the encryption attribute set $\gamma$.
On the other hand, the key size of a decryption key remains the same of the segment tree representation.
Indeed, the $\mathcal{T}_\mathcal{R}$ subtree of a decryption key has the same structure of the segment tree representation, but each attribute included in the $\mathcal{T}_\mathcal{R}$ subtree is one among the $\varepsilon$ replicas. %is chosen among the $\epsilon$ alternatives.
We will refer to this implementation as the \emph{attribute-pool representation}.

To implement this representation, the TTP maintains a field named $num_{i_\omega}$ for each replica in $\mathcal{U}_\mathcal{R}$, which represents the number of non-revoked keys which are using that replica.
%For each attribute in $\mathcal{U}_\mathcal{R}$, the TTP maintains a field named $num_i$ initialized to zero.
The field is initialized to zero. Its value is incremented when the TTP issues a new decryption key whose access policy attribute set contains the attribute $i_\omega$.
It is decreased in case of key revocation.
When the TTP executes the $\mathrm{KeyGen}$ primitive, for each attribute which forms the $\mathcal{T}_\mathcal{R}$ subtree, it chooses the least frequently used replica of the attribute, i.e., $\min_\omega(num_{i_\omega})$.
%This guarantees a differentiation among the keys and
This representation reduces the number of affected users due to a key revocation.
%chooses one among the $\varepsilon$ alternatives for each attribute $i$ which constitutes the $\mathcal{T}_\mathcal{R}$ subtree of the key access policy. % of the related access policy attribute set. (però solo la parte del subtree R...)
%Specifically, it selects the least frequently used alternative of an attribute, i.e., $\min_\omega(num_{i_\omega})$, in order to differentiate the decryption keys.
For example, two users who monitor exactly the same road segments may share no attributes at all if they have different replicas of each attribute in their decryption keys.
In such a case, if one of the two users is revoked, the other will not be an affected user.

\subsection{Time Representation: Universe Subset $\mathcal{U}_\mathcal{X}$}

%We represent time to implement the key validity period mechanism, i.e., to avoid that a key decrypts
In our system, the level of granularity of the time is one day.
We define the \emph{maximum system lifetime} as the maximum number of days of operation of the system.
%, which includes the range of all the possible days of our system, is represented by means of a segment tree.
%The \emph{system lifetime}, which includes the range of all the possible days of our system, is represented by means of a segment tree.
We represent the days from 1 to the maximum system lifetime by means of a segment tree.
%ABE-sealed keys are labelled with a point representation set ($\gamma_\mathcal{X}^{(j)}$), which represents the data production date.
The $\gamma_\mathcal{X}^{(j)}$ set of each ABE-sealed key is formed by the point representation set of the data production date. %, and contains $\mathcal{O}(\log \rho)$ attributes.
%the data production date,
%while key access policies embed the validity period of the decryption key.
%Therefore, ABE-sealed keys are labelled with a point representation set ($\gamma_\mathcal{X}^{(j)}$), which represents its data production date.
On the other hand, the validity period of a decryption key is obtained through a point-in-interval access policy, which forms the $\mathcal{T}_\mathcal{X}$ subtree.
%Note that both the point and the interval representation sets includes $\mathcal{O}(\log \rho)$  attributes.
Our system provides the planned expiration of the decryption keys. % and guarantees th
%In addition, it guarantees that a decryption key is capable of decrypting only the ABE-sealed keys generated during the validity period of such key.
When a decryption key expires, it is not capable of decrypting new ABE-sealed keys anymore.
From the CSS point of view, key expiration is more lightweight than key revocation because no proxy re-encryption is needed.
However, an expired key can still be used to decrypt past ABE-sealed keys produced within the validity period of the key.
If we want to avoid this, we need instead to revoke the key.

Note that applying attribute-pool representation to the universe subset $\mathcal{U}_\mathcal{X}$ does not improve the performance of the revocation since these attributes are never updated.

\section{Experimental Evaluation}\label{experimentalevaluation}

To test our scheme, we used a street network representing the city of Pisa\footnote{Lat: $43.7052$--$43.7264$, Lon: $10.3867$--$10.4266$.} obtained from OpenStreetMap.
%Then, we built three versions of the system by using the different universe representations explained in Section \ref{keyas}.
We assumed a maximum system lifetime of $100$ years. % and allocated a segment tree of proper size to represent the universe subset $\mathcal{U}_\mathcal{X}$.
%We considered the worst case for the cardinality of the $\gamma_\mathcal{X}^{(j)}$ attribute set, which is $17$.
Then, we considered a dataset of $300$ users with $365$-day subscriptions.
%Therefore, by using such segment tree, w
In our simulation, every user subscribes to a \emph{route} composed by consecutive road segments.
A route is characterized by a length (\emph{route length}, $L$), which is the line-of-sight distance between its source point and its destination point.
We chose a source point at random within the map and a destination point at random at a distance $L$ from the source point.
%For each user we created three equivalent versions of his/her decryption key. % by using the different representations of the access policy explained in Section \ref{keyas}.
%Specifically, each version has a different representation of the $\mathcal{T}_\mathcal{R}$ subtree, while they all share the same representation of the $\mathcal{T}_\mathcal{X}$ subtree.
%Specifically, we used the same $\mathcal{X}$ subtree for all the versions, but we used different representations for the $\mathcal{R}$ subtree.
%We call $K_B$ the decryption key whose $\mathcal{T}_\mathcal{R}$ subtree is created through the basic representation,
%$K_S$ the decryption key whose $\mathcal{T}_\mathcal{R}$ subtree is created through the segment tree representation, and
%$K_A$ the decryption key whose $\mathcal{T}_\mathcal{R}$ subtree is created through the attribute-pool representation.
%
%We assumed the system lifetime equal to $100$ years and allocated a segment tree of proper size.
%For the cardinality of the $\gamma_\mathcal{X}$ attribute set we considered the worst case, which is $17$.
%Then, we considered one-year subscriptions.
%%Therefore, by using such segment tree, w
%The average cardinality of the $\mathcal{X}$ subtree, i.e., of an interval representation set of $365$ days, resulted $8.50 \pm 0.12$ ($95\%$ confidence interval).
%%This value corresponds to the average number of attributes of an $\mathcal{X}$ subtree.
%
%We evaluated the average key size and the average percentage of affected users due to a key revocation for the different keys representations.
Within each scenario we tested, we fixed a value for the route length which is the same for all the users.
We tested the system with route lengths of \mbox{$500\,\mathrm{m}$}, \mbox{$1000\,\mathrm{m}$}, and \mbox{$2000\,\mathrm{m}$}.
We used a KP-ABE implementation written in C~\cite{zheng2014kpabe} which realizes the four KP-ABE primitives described in Section \ref{Sec:KPABE} with 80-bit security.

\begin{figure*}[t]
\centering
\subfloat[Average key size wrt route length.]{%
\includegraphics[width=.32\textwidth]{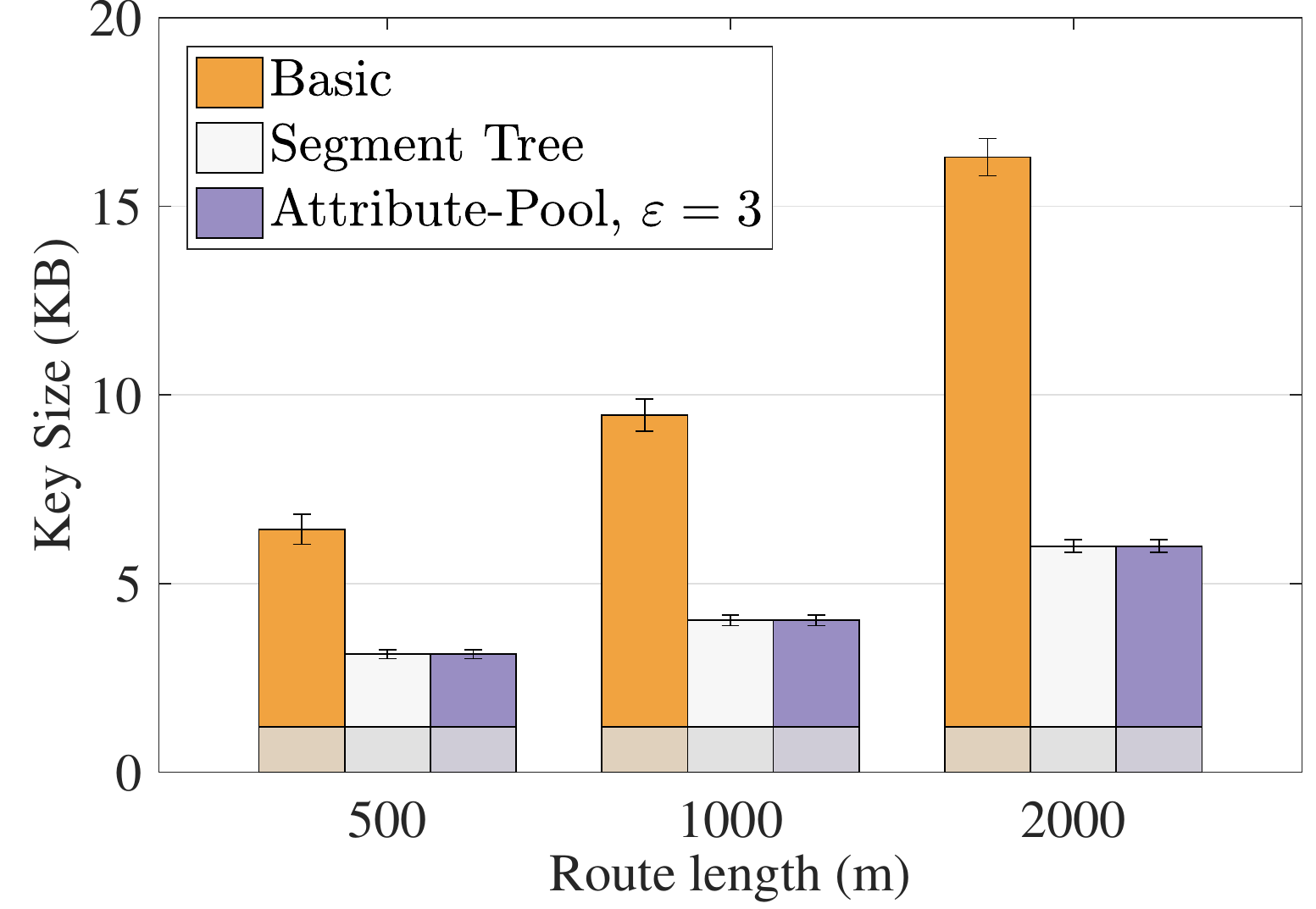}%
\label{fig:plotKeySize}}
\hfil
\subfloat[Average percentage of affected users due to a single key revocation wrt route length.]{%
\includegraphics[width=.32\textwidth]{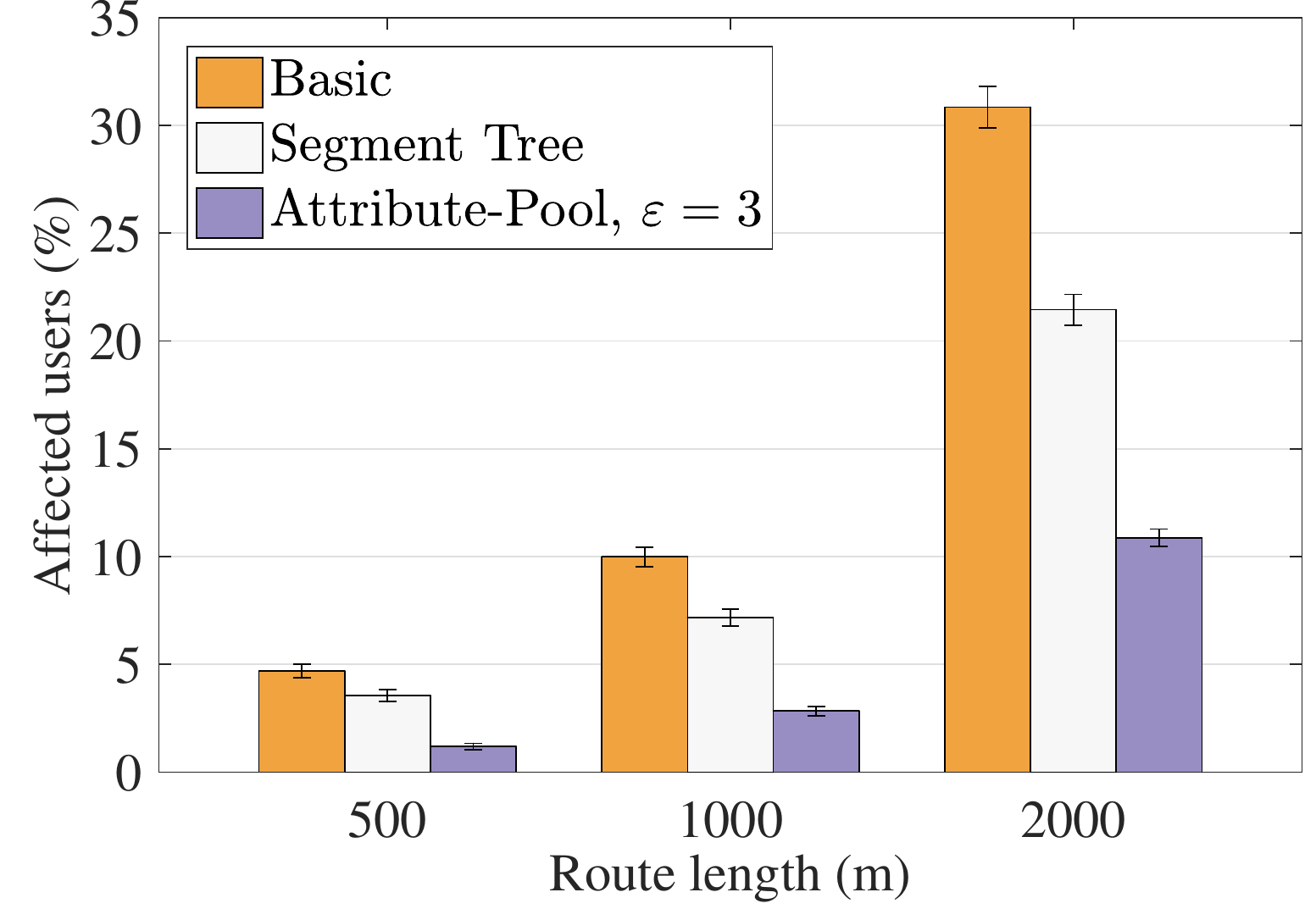}%
\label{fig:revPerc}}
\hfil
\subfloat[Average percentage of affected users due to a single key revocation wrt number of replicas in the attribute-pool representation. $L= 2000\,\mathrm{m}$.]{%
\includegraphics[width=.32\textwidth]{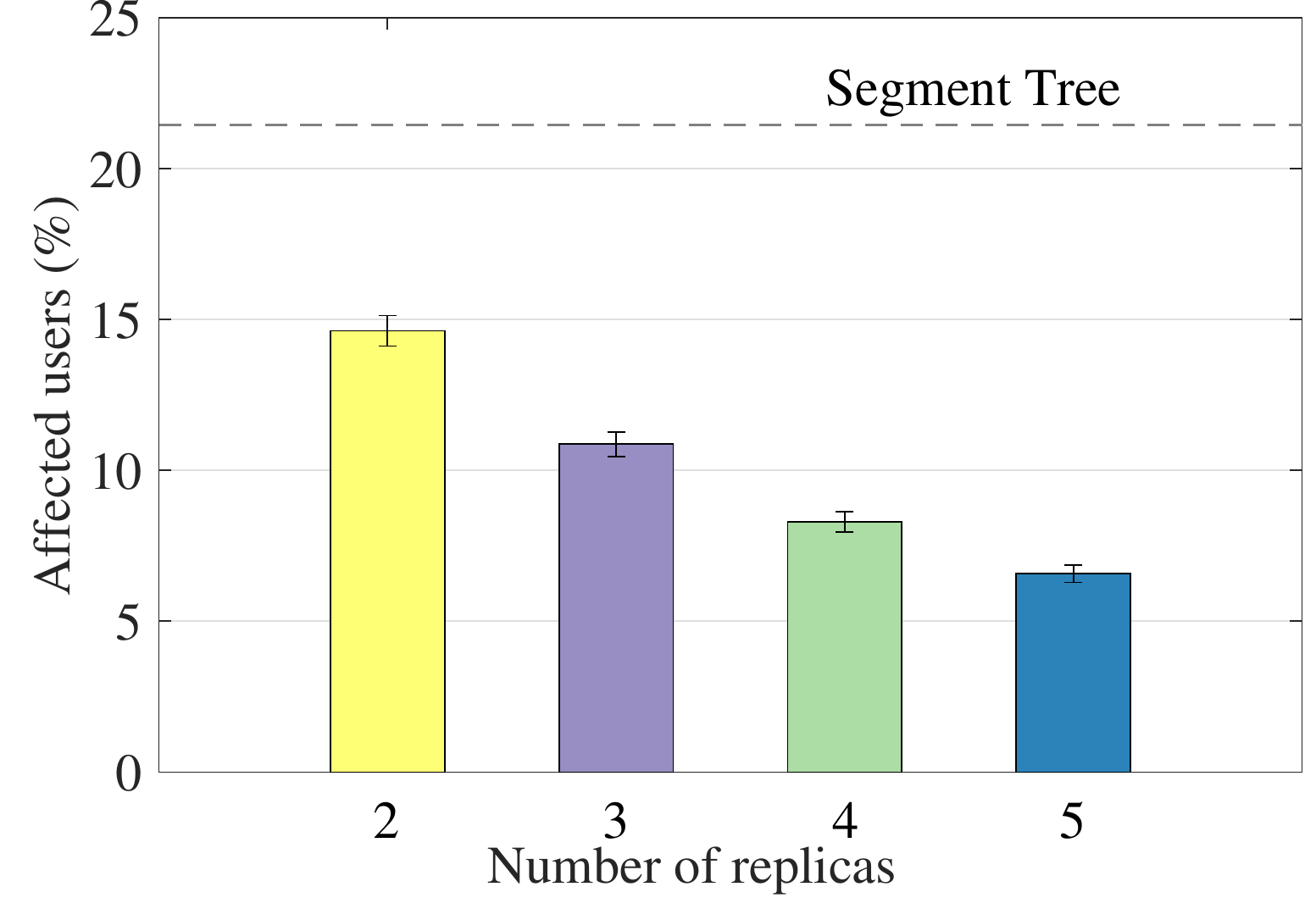}%
\label{fig:compRevEq}}
\caption{Performance evaluation. All the plots show $95\%$ confidence intervals.}
\label{fig:test}
\end{figure*}

In Fig. \ref{fig:plotKeySize} we show a comparison between the average key sizes for the three representations, with respect to the route length.
%%%%For the attribute-pool representation we chose $\varepsilon=3$. %a number of alternative attributes equal to three ($\varepsilon=3$).
%The plot exhibits the results for different values of $L$.
%The reported key sizes include the attributes in the $\mathcal{T}_\mathcal{X}$ subtree, which are represented in the lower part of the bars.
The lower part of the bars shows the portion of the key size concerning the $\mathcal{T}_\mathcal{X}$ subtree.
%The average cardinality of a $365$-day interval representation set,  i.e., of the $\mathcal{T}_\mathcal{X}$ subtree of a decryption key, resulted $8.50 \pm 0.12$ ($95\%$ confidence interval).
%This value corresponds to the average number of attributes of an $\mathcal{X}$ subtree.
%We considered a dataset of $300$ users.
%
%Whatever the route length, the keys in the segment tree representation are always smaller than those in the basic representation, as we expected.
%Likewise, the attribute-pool representation shows the same average key size of the segment tree representation since the access policies have the same structure.
%
%%%%For greater values of $L$, the gain in size reduction with respect to the basic representation is even more relevant. %key size of $K_S$
%The average key size of $K_S$ and $K_A$ is $62.73\%$ less than $K_B$ considering the maximum route length value tested.
%Nevertheless, if we want to know how the representation leverages this outcome, we should look at the cardinality of the $\mathcal{R}$ subtrees only.
%Here, the segment tree and the attribute-pool representations sizes are reduced of $68.33\%$ with regards to the size using the basic representation when $L = 2000$\,m.
%%%%If we consider the impact of the $\mathcal{T}_\mathcal{R}$ subtrees only, the average key size of the segment tree and attribute-pool representations is reduced of $68.33\%$ with respect to the basic representation.
By using any representation, a user's device, e.g., a smartphone, can easily store a decryption key of a few kilobytes.
The CSS stores only the decryption key components of the $\mathcal{T}_\mathcal{R}$ for all the users, and the total size for our dataset of $300$ keys is about $1.4\,\mathrm{MB}$ using either the segment tree or the attribute-pool representation with route length of $2000\,\mathrm{m}$. %size
The key size of the segment tree and the attribute-pool representations are about a third compared to the basic representation.
Nevertheless, the key sizes are very small and all the representations can be used from the point of view of both the CSS and the user.

%The average key size of all the representations is very small and resulted at most about $17\,\mathrm{KB}$ (basic representation, route length = $2000\,\mathrm{m}$).
%This means that all the representations are suitable in terms of storage for users devices
%Thus, users devices, e.g., smartphones, can store them 

%\begin{figure}[htbp]
%\centering
%\includegraphics[width=.8\columnwidth]{keySize_ci.eps}
%\caption{Key size comparison of basic, segment tree, and attribute-pool ($\varepsilon=3$) representations for different values of route length. The plot shows $95\%$ confidence intervals.}
%\label{fig:plotKeySize}
%\end{figure}

By using the same dataset, we revoked each user in turn and measured how many other users were affected.
%we also extracted the average percentage of affected users due to a key revocation.
%To do so, we revoked each key in turn and checked if the other keys shared any attribute --in the $\mathcal{T}_\mathcal{R}$ subtree-- with the revoked one.
In Fig. \ref{fig:revPerc} we show the average percentage of affected users by a single revocation.
%\begin{figure}[htbp]
%\centering
%\includegraphics[width=.8\columnwidth]{revPerc_ci.eps}
%\caption{Percentages of affected users due to a key revocation. Comparison of basic, segment tree, and attribute-pool ($\varepsilon=3$) representations for different values of route length.}
%\label{fig:revPerc}
%\end{figure}
As we expected, the segment tree and the attribute-pool representations show less affected users than the basic representation.
This holds for all the values of the route length tested.
%The number of collisions of the segment tree representation might be at most equal to the one obtained with the basic representation, but our simulation results show that in our scenarios this value is always lower.
%Finally, we compared the affected users experienced by segment tree representation with the ones experienced by the attribute-pool representation, for different values of $\varepsilon$.
Fig. \ref{fig:compRevEq} shows the affected users with respect to the number of alternatives in the attribute-pool representation.
%\begin{figure}[t]
%	\centering
%	\subfloat[Average key size wrt route length.]{%
%		\includegraphics[width=.75\columnwidth]{keySize_ci.eps}%
%		\label{fig:plotKeySize}}
%	%\hfil
%	\\
%	\subfloat[Average percentage of affected users due to a single key revocation wrt route length.]{%
%		\includegraphics[width=.75\columnwidth]{revPerc_ci.eps}%
%		\label{fig:revPerc}}
%	%\hfil
%	\\
%	\subfloat[Average percentage of affected users due to a single key revocation wrt number of replicas in the attribute-pool representation. $L= 2000\,\mathrm{m}$.]{%
%		\includegraphics[width=.75\columnwidth]{compRevEq.eps}%
%		\label{fig:compRevEq}}
%	\caption{Performance evaluation. All the plots show $95\%$ confidence intervals.}
%	\label{fig:test}
%\end{figure}
It is evident from the figure that the affected users can be reduced even further by increasing the parameter $\varepsilon$.
The proposed representations are capable to reduce the number of affected users, and this lightens the load for the CSS due to a key revocation, which is distributed over time thanks to the lazy fashion of proxy re-encryption.
The drawback is that the size of the universe subset $\mathcal{U}_\mathcal{R}$ and the number of attributes of each encryption attribute set $\gamma$ grows with $\varepsilon$. %linearly

In our system, the load on the TTP is mainly determined by the complexity of the $\mathrm{Encrypt}$ primitive, which grows linearly with the cardinality of the encryption attribute set.
The TTP executes such primitive once a day for each device during the seal procedure.
We evaluated the computational load and the required bandwidth on the TTP for the seal procedure.
% either after a key revocation procedure or once a day at midnight.
%Table \ref{tab:cardinality} shows the average cardinality of an encryption attribute set using the attribute-pool representation.
%\begin{table}[!t]
%\caption{Average cardinality of the encryption attributes set $\gamma$.
%The average cardinality of $\gamma_\mathcal{X}$ is $17$.}
%\label{tab:cardinality}
%\centering
%\begin{tabular}{| C | C | C |}
%\hline
%	\varepsilon	&	\gamma_\mathcal{R} &	\gamma	\\
%\hline
%	2 &			8.53 \pm 0.07 &		25.53 	\\
%	3 &			12.80 \pm 0.11 &		29.80	\\
%	4 &			17.07 \pm 0.14 &		34.07 	\\
%	5 &			21.33 \pm 0.18 &		38.33	\\
%\hline
%\end{tabular}
%\end{table}
We used the same street network representing the city of Pisa and a maximum system lifetime of $100$ years.
We assumed $100$ sensing devices deployed randomly on the street network.
We used the attribute-pool representation with $\varepsilon = 5$, which represents the worst case in terms of encryption attributes, and thus the worst case for the computational load and the required bandwidth on the TTP.
We define the \emph{key sealing time} as the time needed to generate all the ABE-sealed keys produced during the daily seal procedure.
We evaluated the key sealing time on a desktop computer equipped with 16GB of RAM, an Intel\textsuperscript{\textregistered} Core\textsuperscript{\texttrademark} i5-6600 CPU, and running Ubuntu 16.04.3 LTS 64-bit operating system.
% the time needed to generate all the ABE-sealed keys produced during the daily seal procedure. % dire che e' un test verosimile dato che e' la TTP che le genera?

From our tests, the average cardinality of an encryption attribute set $\gamma$ is about $39$, and $\gamma_\mathcal{R}$ is about $21$.
%Specifically, $\gamma_\mathcal{X}$ is $17$ as worst case, since we considered a system lifetime of a hundred years. 
%$\gamma_\mathcal{R}$ is $21.33 \pm 0.18$.
We observed that the key sealing time was on average $5.4$ seconds, and the total size of the ABE-sealed keys was about $550\,\mathrm{KB}$. %always less than $9$ minutes.
Hence, the TTP is asked to carry out a task of a few seconds per day and then send a few kilobytes of data to the CSS.

\section{Conclusions}\label{conclusions}

We presented ABE-Cities, an encryption system for urban sensing in a smart city based on Attribute-Based Encryption.
ABE-Cities allows for fine-grained access control over the encrypted data stored in the CSS, and it is secure against multiple adversary models.
%: a honest-but-curious cloud service, external adversaries capable of eavesdropping traffic and compromising sensing devices, and colluding users wanting to gain more authorizations illegally.
Sensing devices perform only lightweight symmetric cryptography operations.
Therefore, ABE-Cities can employ also resource-constrained sensing devices, which makes it suitable for a broader set of IoT applications for smart cities.
Our system also provides mechanisms to enforce planned expiration of decryption keys, as well as their unplanned revocation.
We proposed methods to make the unplanned revocation efficient from the point of view of the cloud service.
We showed by simulations the effectiveness of such improvements.

\bibliographystyle{IEEEtran}
\bibliography{IEEEabrv,bibliography}

\end{document}